\documentclass[aps,prb,twocolumn,groupedaddress,showpacs,superscriptaddress,amssymb,amsmath,longbibliography,]{revtex4-1}
\usepackage{physics}

\usepackage{amsmath}
\usepackage{mathptmx,bm}

\DeclareMathAlphabet{\mathcal}{OMS}{cmsy}{m}{n} 

\usepackage{graphicx}
\usepackage{dcolumn}
\usepackage{tabularx}
\usepackage{epsf}
\usepackage{color}

\usepackage{hyperref}
\hypersetup{breaklinks,colorlinks,linkcolor=blue,citecolor=blue,urlcolor=blue}

\usepackage{epstopdf}%
\usepackage{verbatim}
\usepackage[english]{babel}
\newcommand{\bea}{\begin{eqnarray}}
\newcommand{\eea}{\end{eqnarray}}
\begin{document}

\title{Capturing non-Markovian dynamics with the reaction coordinate method}

\author{Nicholas Anto-Sztrikacs}
\affiliation{Department of Physics, 60 Saint George St., University of Toronto, Toronto, Ontario, Canada M5S 1A7}

\author{Dvira Segal}
\affiliation{Department of Chemistry and Centre for Quantum Information and Quantum Control,
University of Toronto, 80 Saint George St., Toronto, Ontario, M5S 3H6, Canada}
\affiliation{Department of Physics, 60 Saint George St., University of Toronto, Toronto, Ontario, Canada M5S 1A7}
\email{dvira.segal@utoronto.ca}

\date{\today}
\begin{abstract}
The reaction coordinate (RC) technique is emerging as a significant tool in 
the study of quantum dissipative dynamics and quantum thermodynamics.
With the objective to further establish this tool, here we explore to what extent the method can capture non-Markovian dynamics of 
open quantum systems. 
As a case study, we focus on the pure decoherence model of a spin coupled to a harmonic reservoir.
We compare the spin dynamics and measures for non-Markovianity from the exact analytical solution to simulations
based on the RC method at the level of a second order quantum master equation.
We find that the RC method can quantitatively capture non-Markovian effects at strong system-bath coupling and for structured baths. This is rationalized by the fact that the collective RC bath mode, which is made part of the system, maintains system-bath correlations.
Lastly, we apply our RC method and study the spin-boson model in the non-Markovian regime. 
\end{abstract}
\maketitle

\section{Introduction}
\label{Sec:introduction}

Quantum systems are never completely isolated from their surroundings, thus
a build-up of system-bath correlations and the exchange of e.g., particles, energy and
information between the system and their environment is inevitable. 
Theoretical techniques that can faithfully capture these effects are required.
This research area, traditionally developed to treat chemical problems in the condensed phases \cite{Nitzan,Weiss} and light-matter (quantum optics) problems \cite{Breuer} has been recently receiving much attention in relation to the field of quantum thermodynamics and more broadly quantum technologies \cite{Lidar}. 

Proposals to design quantum thermal machines that build on nontrivial quantum effects such as quantum coherences, correlations and quantum statistics are analyzed with a range of open quantum system methods, particularly, quantum master equations (QMEs). Such methods rely on two approximations to make the dynamics tractable: (i) Under the Born approximation one assumes that the system-reservoir coupling strength is weak, thus the dynamics can be solved to second order in the system-reservoir coupling parameter. This assumption results in ignoring the buildup of correlations between the system and its surroundings. (ii) QMEs such as the Redfield equation  rely on the Markov approximation. This assumption concerns the lack of memory in the  dynamics of the system, making it  time-local, essentially ignoring any dynamical back flow from the reservoir to the system  \cite{Breuer,Nitzan,Weiss}. 
Though such second-order QME approaches yield accurate results within their regime of validity, they are quite restrictive and do not convey the correct dynamics when system and reservoirs influence each other's dynamics, as is the case of non-Markovian evolution \cite{Rivas2014,Wiseman,Breuer_2012,deVega2017,Li_2019}. 

Qualitatively, non-Markovian memory effects typically emerge at strong system-bath coupling and when the bath
spectral density function is structured such that specific modes are more strongly coupled to the system.
The unwritten rule is that non-Markovian dynamics demonstrate as 
recoherence effects and a departure from monotonic exponential decay.

\begin{figure}
\centering
\includegraphics[width=.95\columnwidth]{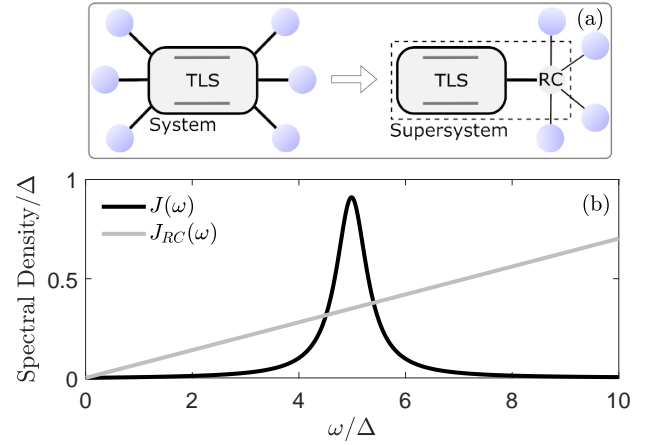} 
\caption{(a) Representation of the RC transformation applied 
onto a two-level system coupled to a bosonic environment.
(b) The Brownian  spectral density function of the bath prior to mapping, $J(\omega)$, and post mapping, 
$J_{RC}(\omega)$, which is Ohmic.
Parameters are $\Omega = 5$, $\gamma = 0.071$, $\lambda = 1.0$, 
 $\Lambda = 1000 \pi$, $\Delta =1$.}
\label{fig:Diagram}
\end{figure}

The reaction coordinate mapping \cite{Nazir18} seems well-suited to describe non-Markovian scenarios.
This method, suggested in Refs. \cite{Burghardt1,Burghardt2,Burghardt3},
redefines the system-environment boundary with the identification and extraction of a 
central (collective) degree of freedom of the environment, termed the reaction coordinate (RC). 
The quantum system is then extended to become a ``supersystem", which  comprises the original, 
pre-mapped quantum system, the RC, and  the interaction of the RC with the original system. For a schematic representation, see  Fig. \ref{fig:Diagram}(a).
%

The advantage of the exact RC mapping becomes clear once we implement
a second-order QME technique, such as the Born-Markov Redfield QME on the supersystem. 
The resulting technique, termed the RC-QME method, is non-perturbative in the original coupling parameter.
Since the method enables cheap computations,
it  has been widely used in studies involving strongly-coupled system-reservoirs.
For example, it was utilized for studying quantum dynamics of impurity models \cite{NazirPRA14,Nazir16,Camille}, thermal transport in nanojunctions \cite{Correa19,Nick2021},  the operation of quantum thermal machines \cite{Strasberg_2016,Newman_2017,Newman_2020}, and transport in electronic systems \cite{GernotF,GernotF2,McConnel_2021}.

%
Earlier open-quantum-system methods were built on similar principles, 
treating non-Markovian baths by e.g. extending the system or by building 
two-tiered environments (primary and secondary modes) and evolving dynamics with
reduced equations of motion in the form of Langevin \cite{Adelman}, Fokker Planck \cite{Tanimura}
or master \cite{Garraway,Tannor} equations; cited works are representative of a very 
rich literature.
More recent works performed chain mapping \cite{chainPlenio,chainDario}  by iteratively adding
bath modes to the system.
Given recent studies of quantum transport and thermodynamics problems with the RC method, 
particularly for addressing system-bath coupling effects 
\cite{Strasberg_2016,Newman_2017,Newman_2020,Nick2021,Camille}, here we focus on this specific formulation 
of nonperturbative open quantum system dynamics.

Due to the explicit  inclusion of environmental degrees of freedom within the collective RC coordinate, 
it is expected that the RC-QME technique  would properly capture non-Markovian effects.
However, as of yet, no detailed study had attempted to address this issue using quantifiable non-Markovianity metrics, which is the objective of our work.

Quantification of non-Markovian dynamics has received ample attention in the last decade with various approaches suggested to define and compute it.
Common measures of non-Markovianity are classified either as divisibility \cite{RHP}, or trace distance quantifiers \cite{BLP}. The former defines a process as being non-Markovian if the quantum map of the open-system time evolution lacks the divisibility criterion stemming from the semi-group property. A trace distance quantifies the distinguishability of quantum states.
Other measures proposed in the literature rely, for instance, on the quantum Fischer information 
\cite{Wang_2010}, correlations \cite{Luo_2012,Fanchini_2014}, or, the volume of the Bloch sphere \cite{Pasterno_2013}, to name a few. Comparative studies of different non-Markovianity metrics were given e.g. in Refs. \cite{Maniscalco2014,Vacchini2014}. 
Beyond definitions and metrics, efficient computational methods have been developed
to evaluate these measures, see for example \cite{He11} , 
including machine learning techniques \cite{Fanchini_2021}.
Most case studies of non-Markovian dynamics have dealt with few 
qubit systems \cite{Fanchini_2021,Vacchini2014} due to the difficulty in
computing metrics of non-Markovianity.
More complex models such as a pure decoherence model with a squeezed bath \cite{He_2019}
or models of coupled qubits \cite{Shen_2017} have also been considered.

In this work, our objective is to assess the capability of the reaction coordinate technique in capturing non-Markovian dynamics.
Towards this goal, we examine the exactly-solvable pure decoherence model, which describes the phenomenon of phase-loss between 
states building a quantum superposition.
Studies of decoherence are central to understanding the quantum-to-classical crossover \cite{Zurek}
and to the development of quantum computing devices \cite{Shor}. 
Decoherence dynamics has been explored in experiments, see e.g. \cite{exp-PD,Bijay,Liu2019}.
Recent proposals suggest to employ it for pure-state thermometry \cite{GooldT21}. 
Other recent studies examined corresponding heat transfer problem in the 
thermal bath \cite{KilgourT,Goold21}.

We quantify non-Markovianity in the pure decoherence model using 
two well-studied methods: (i) The Breuer, Laine and Piilo  (BLP) measure, which concerns trace distance between states
 \cite{BLP}. (ii) The Rivas, Huelga, Plenio (RHP) measure, which probes the complete positivity (CP) divisibility of the dynamical map \cite{RHP}.
By comparing the decoherence dynamics and measures for non-Markovianity from RC-QME simulations to exact expressions, 
we can appraise the RC-QME method, indeed judging it to excellently detail non-Markovian effects in time evolution.

The paper is organized as follows. In Section \ref{Sec:Model}, we introduce the pure decoherence model and the reaction coordinate mapping. The RC mapping and the RC-QME method are presented in Sec. \ref{Sec:Method}. 
Measures for non-Markovianity are discussed in 
Sec. \ref{Sec:Measures}, with simulations presented in Sec. \ref{Sec:Results}. 
We discuss non-Markovianity in the more complex dissipative spin-boson model in  Sec. \ref{Sec:Conclusion},
then  conclude our work.

\section{Pure decoherence model}
\label{Sec:Model}

We begin with the pure decoherence Hamiltonian \cite{Breuer,Weiss}, 
\bea
 \hat H = \frac{\Delta}{2}\hat \sigma_z + \hat \sigma_z \sum_k f_k (\hat c_k^{\dagger} + \hat c_k) + 
\sum_k \nu_k \hat c_k^{\dagger}\hat c_k.
 \label{eq:Hamiltonian_Dephasing}
\eea
Here, $\Delta$ is the energy splitting of the spin, $\hat \sigma_{x,y,z}$ are the Pauli matrices,
$\hat c_{k}^{\dagger}$ ($\hat c_{k}$) are bosonic creation (annihilation) operators of the reservoir 
for a mode of frequency $\nu_{k}$. 
The bath's harmonic oscillators are coupled to the spin polarization with strength $f_{k}$. 
The effects of these interactions are captured by a spectral density function, 
$J(\omega) = \sum_{k}f_{k}^2\delta(\omega-\nu_{k})$.  
Since the system's Hamiltonian and its interaction with the bath commute, the 
system does not exchange energy with the bath.

Our objective is to focus on the non-Markovian dynamics of the spin. We
amplify such non-Markovian evolution by choosing the spectral density function of the environment to be a Brownian function,
peaked around a central mode $\Omega$ with a width parameter $\gamma$ and system-bath coupling strength $\lambda$ \cite{Note},
\bea
J(\omega) = \frac{4 \omega \Omega^2 \lambda^2 \gamma/\pi}{(\omega^2 - \Omega^2)^2 + (2 \gamma \Omega \omega)^2}. 
\label{eq:Brownian}
\eea
Technically, this spectral function arises naturally in the derivation of the reaction coordinate mapping 
when one works backwards,  first defining an Ohmic spectral function for the residual bath
after the mapping \cite{NazirPRA14}.

Following a procedure analogous to that described in Refs. \cite{NazirPRA14,Nick2021}, 
we map the pure decoherence Hamiltonian, Eq. (\ref{eq:Hamiltonian_Dephasing}) 
using the reaction coordinate transformation. 
This mapping consists of preparing a collective coordinate of the environment, 
called the reaction coordinate (RC) and redefining the boundary between the system and the 
reservoir to include this coordinate as part of the system. 
Post mapping, the physical picture consists of a spin coupled to the RC, which in turn couples to a 
residual harmonic reservoir. 
The reaction coordinate is a harmonic-bosonic mode with creation (annihilation) operator $a^{\dagger}$ ($a$).
It is defined such that
\bea
 \lambda(\hat a + \hat a^{\dagger}) = \sum_k f_k (\hat c_k + \hat c_k^{\dagger}).
\eea
This process results in the RC pure decoherence Hamiltonian, 
\bea
\hat H_{RC} = \hat H_{ES}+\hat H_B + \hat H_{ES,B}.
\label{eq:HRC}
\eea
It includes an extended system $\hat H_{ES}$, 
a residual bath $\hat H_B$, and their interaction $\hat H_{ES,B}$,
\bea
\hat H_{ES}&=&
\frac{\Delta}{2}\hat \sigma_z + \lambda \hat \sigma_z (\hat a^{\dagger} + \hat a) + \Omega \hat a^{\dagger}\hat a,
\nonumber\\
\hat H_{B}&=& \sum_{k}\omega_{k}\hat b_{k}^{\dagger}\hat b_{k},
\nonumber\\
\hat H_{ES,B} &=& (\hat a^{\dagger}+\hat a)\sum_k g_{k} (\hat b_{k}^{\dagger}+\hat b_{k}) 
 +  (\hat a^{\dagger} + \hat a)^2\sum_k\frac{g^2_{k}}{\omega_{k}}.
\label{eq:RC_Dephasing_Hamiltonian}
\eea
%
Here, $\hat b_k^{\dagger}$($b_k$) are the creation (annihilation) operators of the residual bath,
$\Omega$ and $\omega_k$ are the frequencies of the RC and modes of the reservoir, respectively.
The energy $\lambda$ corresponds to the coupling strength between the spin and the RC, 
while $g_k$ characterizes the coupling between the RC and the residual environment. 
This system-reservoir coupling (first term in $\hat H_{ES,B}$) is described by the residual environment's spectral density,
 $J_{RC}(\omega)=\sum_{k} g_{k}^2\delta(\omega-\omega_{k})$. 
In correspondence with Eq. (\ref{eq:Brownian}), we set this function to be Ohmic,
\begin{equation}
    J_{RC}(\omega) = \frac{\gamma}{\pi} \omega e^{-\omega/\Lambda}.
        \label{eq:Ohmic}
\end{equation}
Here, $\gamma$ is a dimensionless parameter that captures the coupling strength 
between the residual bath and the extended system, $\Lambda$ is a high frequency cutoff. 
In Fig. \ref{fig:Diagram}(b) we display the spectral density functions of the bath before and after the RC mapping.

By deriving equations of motion for operators in Eq. (\ref{eq:RC_Dephasing_Hamiltonian}), it can be shown that
in the limit of an infinite cutoff,
the corresponding spectral function for the original reservoir is of a Brownian form, 
as depicted in Eq. (\ref{eq:Brownian}). 
We emphasize that the mapping is exact and no approximations have been made so far to obtain the RC pure decoherence Hamiltonian, Eq. (\ref{eq:RC_Dephasing_Hamiltonian}).

As a reminder, the pure decoherence model does not permit 
energy transport between the system and the environment, and it only allows the loss of quantum
coherence in the system. This is because the system's Hamiltonian commutes with the total Hamiltonian:
In the original picture, 
Eq. (\ref{eq:Hamiltonian_Dephasing}),  both system and system-bath coupling operators align with $\hat \sigma_z$.
In contrast, after the RC mapping the extended model obeys $[\hat H_{ES},\hat H_B+\hat H_{ES,B}]\neq 0$. 
%

Given its relative simplicity, it is possible to obtain an exact analytic solution to the decoherring dynamics
of Eq. (\ref{eq:Hamiltonian_Dephasing}) \cite{Breuer,Nitzan,Weiss}. 
Namely, given an initial state for the system (after tracing out the bath), 
$\rho(0) = \begin{pmatrix} \rho_{00} && \rho_{01} \\ \rho_{10} && \rho_{11} \end{pmatrix}$, 
it can be shown that at a later time $t$ the state of the system evolves to
\bea
    \rho(t) = \begin{pmatrix} \rho_{00} && \rho_{01} e^{\Gamma(t) - i\Delta t} \\ \rho_{10}e^{\Gamma(t) + i\Delta t} && \rho_{11} \end{pmatrix}. 
\eea
While populations are constant in time, coherences can show  nontrivial decaying dynamics.
This behavior is captured via the decoherence function (we set Planck's and Boltzmann's constants to
$\hbar\equiv 1$, $k_B\equiv1$),
\bea
\Gamma(t) = -4 \int_0^\infty J(\omega) \coth(\frac{\beta \omega}{2}) \frac{1 - \cos(\omega t)}{\omega^2} d\omega,
\label{eq:Decoherence}
\eea
where $\beta$ is the inverse temperature of the bosonic reservoir. 
%
%
%
In Appendix A we show that
in the long time limit and for nonzero temperature 
the decoherence function reduces to 
\bea
\Gamma(t) \xrightarrow{\text long \,time}  -\frac{16 \gamma \lambda^2}{\Omega^2 \beta} t.
\eea
According to this result, coherences decay exponentially with time, with the decoherence timescale
\bea
\tau_D = \frac{\beta \Omega^2}{16 \gamma \lambda^2}.
\eea
As we show in Appendix A,
this result can be obtained directly from the exact expression, Eq. (\ref{eq:Decoherence}), as well as from
the Redfield equation, which is built on the Born-Markov approximation. 



\section{The RC-QME method}
\label{Sec:Method}

The simplicity of reaction-coordinate simulations stems from the fact that, once the RC is 
extracted from the environment and added to the system, 
we can employ standard Markovian weak-coupling quantum master equation techniques to describe the
dynamics of the supersystem; we use the Redfield equation. 
We refer to this approach, of extracting an RC from the bath then simulating the dynamics of the supersystem
with the Redfield equation as the RC-QME method.
In this section, we briefly describe this method and its implementation on the decoherence model.

We truncate the RC harmonic mode (eigenstates $|n\rangle$) to include $M$ energy levels; 
in simulations, $M$ should be taken sufficiently large to ensure convergence of results. 
The supersystem now includes a spin coupled to an $M$-level system, in turn coupled to a residual heat bath.
%
The Hamiltonian is 
$\hat H^M_{RC} = \hat H_{ES}^M + \hat H_{ES,B}^M + \hat H_B$, corresponding to Eq. (\ref{eq:HRC}), with
%
%
\bea
\nonumber
\hat H_{ES}^M &=& \frac{\Delta}{2}\hat \sigma_z + 
\Omega \sum_{n=0}^{M-1} \left( \frac{1}{2} + n \right ) \ket{n}\bra{n} 
\\ 
&+&  \lambda \hat \sigma_z \sum_{n=1}^{M-1} \sqrt{n} (\ket{n}\bra{n-1} + \ket{n-1}\bra{n}),
\label{eq:RCM}
\\ \nonumber
\hat H_{ES,B}^M &=& \sum_k \sum_{n=1}^{M-1} \sqrt{n} (\ket{n}\bra{n-1} + \ket{n-1}\bra{n})  
g_k \left(\hat b_k + \hat b_k^{\dagger} \right).
\eea
In the last line, we neglected the nontrivial quadratic term: Once performing the Redfield QME, its impact is absorbed in the dissipator \cite{NazirPRA14}.
We can compactify the system-bath interaction  as $\hat H_{ES,B}^M = \hat S^M_{ES} \otimes \hat B$, 
identifying  $\hat S_{ES}^M$ and $\hat B$ from Eq. (\ref{eq:RCM}).
Note that $\hat S_{ES}^M$ is defined in the $2M$-dimensional Hilbert space of the supersystem.

To study dynamics and steady state values, 
we employ the energy-basis Redfield QME. 
As such, we first diagonalize  $\hat H_{ES}^M$ in Eq. (\ref{eq:RCM}) with
a unitary operator $\hat U$, a process that we do numerically. The transformed supersystem operators are
$\hat H^D_{ES} = \hat U^{\dagger} \hat H^M_{ES} \hat U$, 
$\hat S_{ES}^D = \hat U^{\dagger} \hat S^M_{ES} \hat U$.    
%

The Redfield equation is reliable for our model as long as
(i) The coupling energy of the extended system to the residual bath is small. 
In terms of model parameters, this means that $\gamma \Delta \ll T$. 
This assumption is necessary for the second-order perturbative treatment of the master equation to yield 
accurate results. 
(ii) The residual reservoir should be relatively  structureless, supporting a Markov approximation on the supersystem. 
(iii) As an initial condition, the total state is assumed factorized into the supersystem times bath states.
Further, the state of the reservoir is given by a canonical thermal state at temperature $T=1/\beta$.

In the Schr\"odinger picture, the  Redfield time evolution of the extended system is given by
\bea
    \dot{\rho}_{ES}(t) &=& {\cal L}_{red}\rho_{ES}(t)
\nonumber\\
&=&
-i[\hat H_{ES}^D, \rho_{ES}] + D(\rho_{ES}),
    \label{eq:Lindbladian}
\eea
with the Liouvillian ${\cal L}_{red}$ including the unitary dynamics and 
the Redfield dissipator $D$. 
For completeness, these expressions are included in Appendix B.
%
%

We time-evolve the dynamics subject to initial coherences. 
However, since we are not interested in the dynamics of the reaction coordinate itself---but the spin alone---we trace out the RC degree of freedom,  
\bea 
\rho(t) = {\rm Tr}_{RC}\left[e^{{\cal L}_{red} t }\rho_{ES}(0)\right].
\label{eq:RC-QME}
\eea
%
As for the initial condition of the supersystem $\rho_{ES}(0)$  
we assume that the spin and the RC are prepared in a product state,
$\rho_{ES}(0) = \rho(0) \otimes \rho_{RC}(0)$,
with  the RC initially thermalized to the attached reservoir.
Therefore, ignoring the zero point energy (which is not influential here) we write
%
$\rho_{RC}(0) = \frac{e^{-\beta \Omega a^{\dagger}a}}{Z}$, 
%
with $Z = {\rm Tr}_{RC}  \left[ e^{-\beta \Omega a^{\dagger} a} \right]$ 
the partition function of the reaction coordinate. Note that 
in the truncated basis we normalize level populations to unity.
The initial state of the spin is ours to choose.
For the purpose of decoherence dynamics explorations
we arbitrarily set it to be $\rho(0)=\frac{1}{2}(|0\rangle +|1\rangle)(\langle 0| + \langle 1|)$.

The RC-QME time evolution, Eq. (\ref{eq:RC-QME}),
is non-perturbative in $\lambda$, the original system-bath coupling energy.
It is perturbative however in the width parameter $\gamma$,  as we invoke the Born-Markov approximation on the extended model. 
Given that the RC is embedded in this model, we expect non-Markovian effects to be preserved in the dynamics.
To quantify the extent of non-Markovianity captured by the RC-QME method we benchmark it
against exact results based on the decoherence function, Eq. (\ref{eq:Decoherence}). 
We further compare the RC-QME technique to the standard Born-Markov-Redfield quantum master equation  (BMR-QME) on the original 
model without the extraction of the RC. This method handles system-bath couplings $\lambda$ to second order only.


\section{Measures of non-Markovianity}
\label{Sec:Measures}

Unlike classical Markovian dynamics, which is mathematically well-defined 
using classical stochastic processes,  
quantum Markovianity is nontrivial to quantitatively define, and
the literature includes numerous definitions and metrics to quantify it 
\cite{Rivas2014,Breuer,deVega2017,Wiseman,Li_2019}.
Many of these measures can be arranged into two classes, concerning (i) the distinguishability of states
and (ii) divisibility of the dynamical map. 
Underlying such measures are concepts such as
trace-distance, fidelity, negativity of the decay rate, 
system-bath correlations and information flow \cite{Rivas2014}. 
Despite their differences, a unifying aspect concerning measures for non-Markovianity
is the emergence of memory effects within the open quantum system dynamics
often visible with recoherences.

In this work, we do not aim to adjudicate between different quantifiers for
non-Markovianity. With the objective to assess to what extent  the RC-QME 
method can capture non-Markovian dynamics,
we employ two distinct, well-accepted quantifiers for non-Markovianity:
The BLP \citep{BLP} measure on the distinguishability of states, and the  RHP quantifier \citep{RHP} 
on the complete positive (CP) divisibility of the dynamical map.
With extensive simulations and benchmarking against exact results, 
we demonstrate that the RC-QME technique excellently captures 
non-Markovian dynamics in the pure decoherence model.

\subsection{BLP Measure}
\label{sec:BLP}

In their influential work, Breuer, Laine and Piilo presented the so-called BLP measure 
for non-Markovianity \cite{BLP}.
This metric is based on the nonmonotonicity of the trace distance  $D$ between quantum states,
\bea
    D(\rho_1, \rho_2) = \frac{1}{2}||\rho_1-\rho_2|| =\frac{1}{2} \sum_i |\epsilon_i|.
\label{eq:Trace Distance}
\eea
Here, $\rho_1$ and $\rho_2$ are two quantum states,
$\epsilon_i$ are the eigenvalues of the hermitian operator $\rho_1 - \rho_2$. 
The trace distance can be interpreted as a measure for state distinguishability: 
It returns values between $0$  and $1$,
ranging from indistinguishable (0) to distinguishable (1) states. 
A decrease in trace distance is associated with a reduced
ability to distinguish between two states.
In our context, this is associated with a loss of information over the quantum system.
 Conversely, an increase in the trace distance corresponds to a higher ability to distinguish states, 
interpreted as information returning to the open system from the reservoir. This
situation can be attributed to the survival of memory effects in the dynamics.

To quantify non-Markovianity, one adds contributions of the trace distance 
in regions where it is increasing, 
\bea
    {\cal{N}}_{BLP} = \max_{\rho_1,\rho_2} \int_{\sigma > 0} dt \sigma(t,\rho_1,\rho_2).
\label{eq:BLP}
\eea
Here $\sigma(t,\rho_1,\rho_2) = \frac{d}{dt} D(\rho_1(t), \rho_2(t))$. 
where $\rho_{1,2}$ are the initial states and  $\rho_{1,2}(t)$ are time evolved states under a certain dynamical equation. 
Note that the measure is obtained  
after maximization over all pairs of initial states;
it has been shown that the pair of states that maximize the BLP measure, Eq. (\ref{eq:BLP}), are orthogonal.  
Furthermore, it can be shown that the trace distance is contractive for any completely-positive trace-preserving  map. 

Eq. (\ref{eq:BLP}) is not fully transparent, and the required optimization can make it a heavy task to compute \cite{BLP}.
Luckily, for the exactly-solvable decoherence model the states maximizing the increase of the trace distance
can be identified. Thus, an analytic expression can be derived for the BLP measure.
It is given in terms of the decoherence function,
Eq. (\ref{eq:Decoherence}), as \cite{He11,Maniscalco2014} 
\bea
  {\cal{N}}_{BLP} = \sum_j \left[e^{\Gamma(b_j)} - e^{\Gamma(a_j)}\right]. 
\label{eq:NBLP}
\eea
Here, $t\in[a_j, b_j]$ indicates the jth time interval in which the trace distance is increasing.
For a Markovian process, the non-Markovianity measure is zero.


\subsection{RHP Measure}
\label{sec:RHP}

The Rivas, Huelga, Plenio measure for non-Markovianity inquires on the 
CP-divisibility property of the dynamical map \cite{RHP}.
Considering a QME of Lindblad form with a set of time-dependent rates $\gamma_k(t)$,
divisibility, and thus Markovianity is identified if the rates are positive throughout, $\gamma_k(t)\geq0$.
Namely, according to the RHP measure, 
a dynamical process is Markovian if the dynamical map can be split (factorized) at all times. 
A process evolves under non-Markovian dynamics if there is a lack of divisibility. 
Concretely for the pure decoherence model
the RHP measure is given by \cite{Maniscalco2014}  
\bea
    {\cal N}_{RHP} = 
   \sum_j \left[\Gamma(b_j) - \Gamma(a_j)\right],
\label{eq:RHP}
\eea
%
where similarly to before, $t\in[a_j,b_j]$ is the jth interval where nonmonotonic decay takes place.
In contrast to the BLP measure, 
this quantifier does not require optimization, therefore,  it is generally easier to compute than the BLP.

We evaluate Eqs. (\ref{eq:NBLP}) and (\ref{eq:RHP}) in two ways:
(i) By using the exact analytic form, Eq. (\ref{eq:Decoherence}).
(ii) Approximately, by simulating the dynamics
with the RC-QME method as described in Sec. \ref{Sec:Method}. 
In the latter case, we time-evolve 
an initial state of the system to long enough times according to Eq. (\ref{eq:RC-QME}),
then evaluate the decoherence function from the 
ratio  $e^{\Gamma_{RC}(t)}\equiv|\rho_{01}(t)/\rho_{01}(0)|$.
\section{Results}
\label{Sec:Results}

We study the dynamics of the decoherence model, Eq. (\ref{eq:Hamiltonian_Dephasing}),
with (i) the exact expression (Exact) Eq. (\ref{eq:Decoherence}), (ii)
the reaction-coordinate method (RC-QME), and (iii) the Born-Markov Redfield approach without the RC mapping (BMR-QME). 
We emphasize that the BMR-QME is perturbative in the coupling energy $\lambda$, while the RC-QME treats $\lambda$ to all orders, and is instead perturbative in the width parameter, $\gamma$.
As such, for weak and intermediate couplings the RC-QME very well performs against exact results.
We further illustrate its challenges  at strong couplings:
Once $\lambda\gtrsim\Omega$, 
the RC-QME method poorly converges with the number of levels, $M$.
Beyond studies of dynamics, we use the BLP and the RHP metrics to quantify non-Markovianity in the pure decoherence model comparing
results from the exact decoherence function to RC-QME simulations.

\begin{figure}
\centering
\includegraphics[width=\columnwidth]{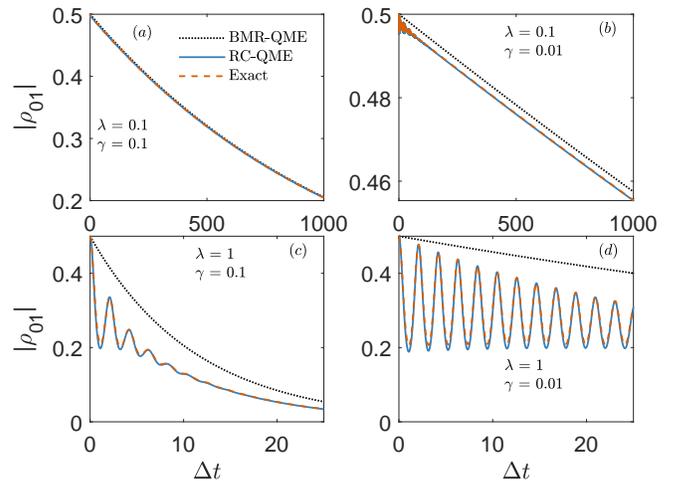}
\caption{Decoherence dynamics with bath structuring (small $\gamma$) 
and enhanced coupling (large $\lambda$).
We display coherences $|\rho_{01}|$ starting from $\rho_{01}(t=0)=1/2$
as a function of the dimensionless
time $\Delta t$
using the Redfield Born-Markov QME (dotted)
the RC-QME method (full), and the exact analytical solution (dashed). 
Parameters are $\Delta = 1$, $\Omega = 3$, $T = 0.5$,
$\Lambda = 1000 \pi $ and $M = 8$. Note, all parameters are given in units of $\Delta$. 
}
\label{fig:Coherence}
\end{figure}

\begin{figure}
\centering
\includegraphics[width=\columnwidth]{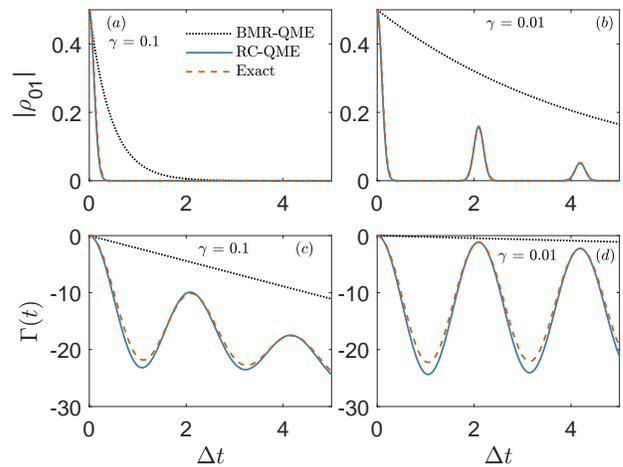} 
\caption{Recoherence behavior at strong coupling, $\lambda = 5$.
Starting from $\rho_{01}=1/2$ we present
(a)-(b)  the dynamics of  $|\rho_{01}(t)|$ and 
(c)-(d) the decoherence function.
Parameters are the same as in Fig. \ref{fig:Coherence}
with the BMR-QME (dotted), the RC-QME with $M=50$, (full), and the exact analytical solution (dashed).}
\label{fig:Coherence_lambda_5}
\end{figure}

\begin{figure}
\centering
\includegraphics[width=\columnwidth]{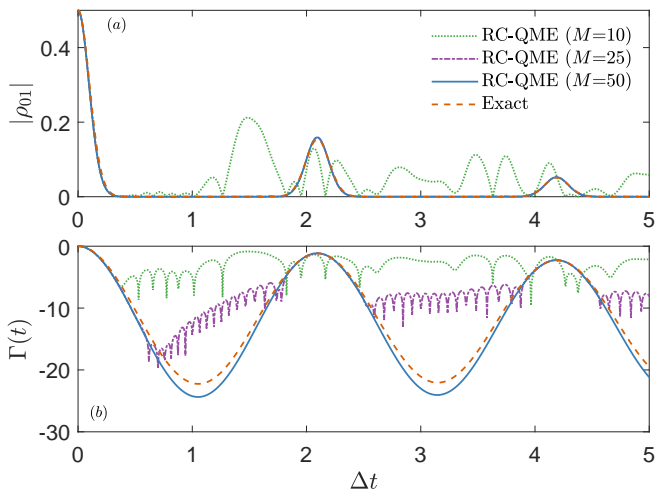}
\caption{Convergence of the RC-QME simulations, Fig. \ref{fig:Coherence_lambda_5} (a) and (c), 
with respect to $M$ at strong coupling, $\lambda=5$.
We present the (a) decoherence dynamics and (b) decoherence function for
$M=10$ (dotted), 25 (dashed-dotted), 50 (full), and compare those to exact results (dashed).
%
}
\label{fig:Convergence_lambda_5}
\end{figure}


\subsection{Decoherence dynamics}
\label{sec:results-dyn}

We examine the role of the system-bath coupling $\lambda$ 
and the bath structuring parameter $\gamma$ on  $\rho_{01}(t)$
in Fig. \ref{fig:Coherence}.
By comparing  RC-QME simulations (full) to the exact expression (dashed)
we conclude that the RC-QME accurately captures 
the  dynamics at both short and long times and 
from small to large $\gamma$ and $\lambda$. 
In contrast, the BMR-QME method (dotted)  fails to follow the correct dynamics of 
$\rho_{01}(t)$ at large $\lambda$ and small
$\gamma$.

Since the BMR-QME method only describes Markovian dynamics
with an exponential decay of coherences (see Appendix A), 
we conclude, qualitatively at this point,
that the RC-QME method properly describes non-Markovian effects, roughly identified here by recoherences.
%
Next, we discuss in more details the role of $\lambda$ and $\gamma$ in the decoherence dynamics.

{\it Structuring of the bath.}
Small width parameter $\gamma$ corresponds to a sharp spectral density function of the thermal bath,
which is commonly associated with the emergence of non-Markovian dynamics.
Indeed, as we reduce $\gamma$ 
we observe deviations from the Markovian-exponential trend,
compare Fig. \ref{fig:Coherence}(a)  to Fig. \ref{fig:Coherence}(b)
and Fig. \ref{fig:Coherence}(c)  to Fig. \ref{fig:Coherence}(d).
Recoherence dynamics is most pronounced
in panel (d) at strong coupling and for small width $\gamma$.

{\it System-bath coupling.} 
We pinpoint the impact of $\lambda$ on the decoherence timescale by
comparing Fig. \ref{fig:Coherence}(a) to Fig. \ref{fig:Coherence}(c).
As expected, this timescale is shorter at strong coupling
in accord with Eq. (\ref{eq:Decoherence}), $\Gamma(t) \propto \lambda^2$. 
The same conclusion is arrived at when comparing Fig. \ref{fig:Coherence}(b) to Fig. \ref{fig:Coherence}(d).
Note on the different timescales presented in these different panels:
The decoherence characteristic time is $\tau_D=  \frac{\beta \Omega^2}{16 \gamma \lambda^2}$,
resulting in $\tau_D\approx 1.13\times 10^{4}$ and $1.13\times 10^{2}$ in 
Fig. \ref{fig:Coherence}(b) to Fig. \ref{fig:Coherence}(d), respectively. Thus, only at around those 
timescales we expect to see significant suppression of coherences.


%
%

We focus on the strong-coupling behavior in Fig. \ref{fig:Coherence_lambda_5}
and present results
in both the Markovian, large $\gamma$ region (Fig. \ref{fig:Coherence_lambda_5}(a) and Fig. \ref{fig:Coherence_lambda_5}(c)) and non-Markovian (small $\gamma$)
regime (Fig. \ref{fig:Coherence_lambda_5}(b) and Fig. \ref{fig:Coherence_lambda_5}(d)). 
We note on the accuracy of the RC-QME method in capturing decoherence dynamics, as
compared to exact results,  in contrast to the complete failure of the BMR-QME method.
This agreement is notable given dramatic recoherence dynamics observed at small $\gamma$ 
in Fig. \ref{fig:Coherence_lambda_5}(b). 

When inspecting the decoherence function in Fig. \ref{fig:Coherence_lambda_5} (c),
we observe oscillatory dynamics, the expected signature of non-Markovianity.
Yet, this rich dynamics does not show up in Fig. \ref{fig:Coherence_lambda_5} (a),
where the magnitude of coherences appears to be monotonically decreasing with no features.
However, we note that the scale of the y-axis in Fig. \ref{fig:Coherence_lambda_5} (c) 
is rather small; recall that $|\rho_{01}(t)| =  |\rho_{01}(0)|e^{\Gamma(t)}$,
thus having little impact on the magnitude of coherences.


{\it Convergence.}
RC-QME simulations should be converged with respect to $M$, 
the number of levels representing the RC harmonic oscillator. 
Simulations in Fig. \ref{fig:Coherence} at small to intermediate $\lambda$
excellently converge with $M=8$.
In contrast, converging the 
large $\lambda$ dynamics  presented in Fig.  \ref{fig:Coherence_lambda_5}
require a large number of RC levels, up to $M=50$.
Thus, the RC-QME method  becomes impractical to use at strong coupling, $\lambda\gtrsim\Omega$ 
and small $\gamma$, as it scales unfavorably with $M$.

To illustrate the difficulty in converging RC-QME simulations at large $\lambda$,
in Fig. \ref{fig:Convergence_lambda_5} we display the decoherence dynamics
for three different $M$ values: 10, 25, and 50. 
Comparing their behavior  in panel (a) 
we notice that $M = 10$ completely misses the dynamics by many orders of magnitude.
Further, according to Fig. \ref{fig:Convergence_lambda_5}(a),   $M=25$ 
seemingly agrees with exact results. However, a careful inspection of the
decoherence function itself in panel (b) reveals significant deviations.
It is only at large $M$, here at $M=50$, that results converge and further reasonably 
agree with exact results.

Interestingly,  Fig.  \ref{fig:Convergence_lambda_5} 
reveals  another aspect of the RC-QME method: It tends to overestimate
the effect of non-Markovianity. 
Staying with a qualitative picture for now, we identify non-Markovianity in our model
with the departure from an exponential decay and appearance of recoherences \cite{Garmon1,Garmon2}.
Fig.  \ref{fig:Convergence_lambda_5}(b)
reveals that the RC technique displays deeper troughs than exact results. This
points to the fact that the inclusion of one specific 
RC mode into the system may lead to an overestimation of non-Markovian features.



\begin{figure*}[htp]
 \centering
 \includegraphics[width=1.4\columnwidth]{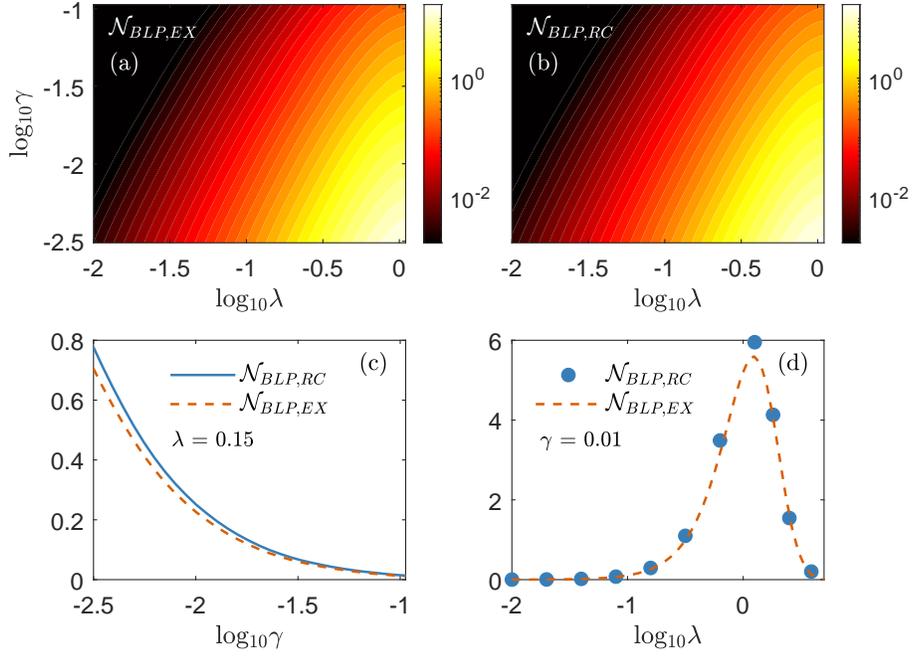} 
\caption{The BLP measure of non-Markovianity as a function of the coupling energy $\lambda$
and spectral density width $\gamma$.
We present contour maps based on (a) the exact dynamics and (b) RC-QME simulations.
We further show
the BLP measure (c) as a function of $\gamma$ at weak coupling (we used $M=8$) and
(d) as a function of $\lambda$ for small $\gamma$ (with $M=50$)
using the exact expression (dashed) and RC-QME simulations (full or circles).
Parameters are the same as in Fig. \ref{fig:Coherence}.
}
\label{fig:BLP_Measure}
\end{figure*}

\begin{figure*}[htp]
 \centering
 \includegraphics[width=1.4\columnwidth]{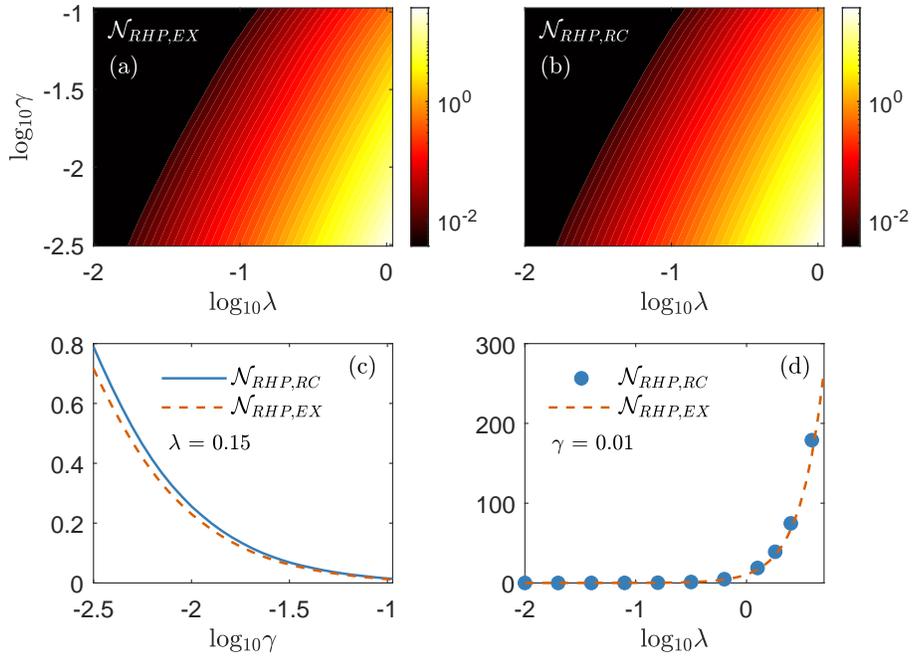} 
\caption{The RHP measure of non-Markovianity  as a function of the coupling energy $\lambda$
and spectral density width $\gamma$.
We present contour maps based on (a) the exact dynamics and (b) RC-QME simulations.
We further show
the RHP measure (c) as a function of $\gamma$ at weak coupling (we used $M=8$) and
(d) as a function of $\lambda$ for small $\gamma$ (with $M=50$)
using the exact expression (dashed) and RC-QME simulations (full or circles).
Parameters are the same as in Fig. \ref{fig:Coherence}.
}
 \label{fig:RHP_Measure}
\end{figure*}
\subsection{Quantifying non-Markovianity}



We quantify the degree of non-Markovianity in the dynamics using the different measures focusing on the performance of the RC-QME method compared to the exact expression.
We adopt the BLP measure for non-Markovianity in Fig. \ref{fig:BLP_Measure} 
and the RHP quantifier in Fig. \ref{fig:RHP_Measure}, examining their 
dependence on the system-bath coupling $\lambda$ and the spectral width $\gamma$.
Each data point in Figs. \ref{fig:BLP_Measure} and \ref{fig:RHP_Measure}
was obtained by processing long time traces of decoherence dynamics using the quantifiers,
 Eqs. (\ref{eq:BLP}) and (\ref{eq:RHP}).
Time traces extend
from initial conditions at $t_i$ to a long time $t_f$
at which recoherences become insignificant.
These figures thus compile significant data; 
while convergence was rapid for small $\lambda$ 
(Fig. \ref{fig:BLP_Measure}(c) and Fig. \ref{fig:RHP_Measure}(c)).
 it was expensive to converge results at large $\lambda$
thus we only show few data points in Fig. \ref{fig:BLP_Measure}(d) and Fig. \ref{fig:RHP_Measure}(d).

In Fig. \ref{fig:BLP_Measure}, we show 
contour plots obtained from the exact solution, ${\cal N}_{BLP,EX}$, in panel (a),
and the RC method  ${\cal N}_{BLP,RC}$, in panel (b).
We note on the excellent agreement between the maps. 
As an example, we plot in Fig. \ref{fig:BLP_Measure}(c) the BLP measure as a function of $\gamma$ at small coupling,
observing its monotonic decay with increased $\gamma$.
This is to be expected, as sharply peaked spectral functions are usually associated with  a longer memory time for the system, 
which in turn leads to non-Markovianity. 
Following this, in Fig. \ref{fig:BLP_Measure}(d) we study the behavior of the BLP measure
with $\lambda$ for a narrow spectral function, 
pushing $\lambda$ to large values beyond what is included in the contour map.
Here, we observe a nonmonotonic behavior: the BLP measure first rises with $\lambda$ at weak coupling,
then it rapidly decays at large $\lambda$.


We examine non-Markovianity with the RHP measure in Fig. \ref{fig:RHP_Measure}, again
as a function of the system-bath coupling $\lambda$ and bath width parameter $\gamma$.
The comparison of contour maps generated from the exact solution (a) and the RC-QME method (b) 
confirms that the RC method  correctly captures non-Markovianity as measured by the RHP method. 
We display the $\gamma$ dependence in Fig. \ref{fig:RHP_Measure}(c), and the $\lambda$ dependence in Fig. \ref{fig:RHP_Measure}(d).
In contrast to the BLP measure, the RHP metric
quickly increases at large coupling without showing a turnover behaviour.
These contrasting trends were also observed and discussed in Ref. \cite{Vacchini2014,Maniscalco2014}.  

Altogether, our main observations are:
(i) The RC-QME method {\it quantitatively} captures non-Markovianity in the dynamics, including nontrivial trends
as displayed in Fig. \ref{fig:BLP_Measure}(d).
(ii) The RC-QME may overestimate the extent of non-Markovianity in the dynamics, presumably due
to the emphasis of a single-collective bath degree of freedom in the system's dynamics.

\section{Discussion and Summary}
\label{Sec:Conclusion}

In this work, we addressed the potential of the reaction-coordinate QME method 
in capturing non-Markovian dynamics.
The advantage of the RC-QME method in this respect is, that by adding a collective bath mode
to the system, one explicitly captures its dynamics, including correlations 
that develop in time between the bath with system's degrees of freedom. 

As a benchmark, we used the exactly-solvable model of pure decoherence in a single qubit. 
By observing the dynamics and quantifying it with two different measures for non-Markovianity, we
showed that the RC-QME method excellently captured such effects in the difficult parameter regimes
of strong system-bath coupling (large $\lambda$) and highly-structured baths (small $\gamma)$.

We used two measures for non-Markovianity, the BLP, which tests states' distinguishability 
and the RHP, which inquires on the CP-divisibility of the dynamical map. Both measures
indicate that  non-Markovian effects  are enhanced
as we reduce the width of the spectral function of the bath.
This result, observed with exact expressions, agrees with common knowledge.

What is important to note is that, very favorably, the RC method becomes 
{\it more accurate} for highly-structured baths.
This is because this scenario precisely fits the principles of the RC mapping, 
of extracting a prominent-collective degree of freedom from the bath 
and adding it to the system.
The capability of the RC-QME method to describe non-Markovian effects is particularly 
notable since powerful numerically exact techniques such as influence functional path integral methods \cite{Makri1,Makri2,Makri20a,Makri20b,Keeling}
face challenges converging for narrow spectral functions.

As for strong coupling effects, we found that the RC-QME becomes increasingly difficult to converge
as we increase $\lambda$, posing a challenge in computations. 
Nevertheless, with increased computational work we found
that the RC-QME method accurately described non-Markovian effects even in the strong coupling limit.


The RC mapping as described in this work was done analytically with the
conversion of a Brownian spectral function to an Ohmic one.
While the Brownian model,  and the related Debye model had 
found applications such as in studies of exciton dynamics in 
the Fenna-Matthews-Olson complex \cite{Thorwart11}, it is important to remember
that the RC mapping can be performed numerically on more general spectral functions, 
as well as iteratively,
to continue and extract primary modes from the bath \cite{Nazir18}.

\begin{figure}
 \centering
\includegraphics[width=\columnwidth]{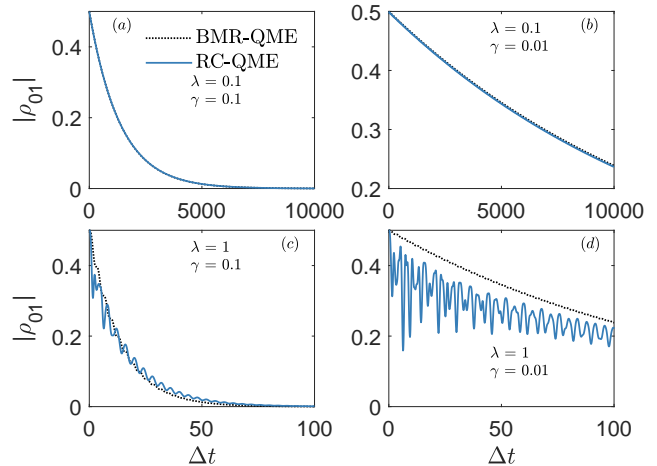}
 \caption{Coherence dynamics of the spin-boson model
for the initial state  $\rho=|-\rangle\langle-|$ with $|-\rangle= (|0\rangle - |1\rangle)/\sqrt{2}$.
Calculations were done with the Born-Markov Redfield equation (dotted line)
and the RC-QME method (full), for details see Appendix C.
Parameters are the same as in Fig. \ref{fig:Coherence}.}
\label{fig:SB_coherence}
\end{figure}

With the goal to gauge non-Markovian effects in the RC-QME method, 
we focused our efforts here on the exactly-solvable pure decoherence model, Eq. (\ref{eq:Hamiltonian_Dephasing}).
It is interesting to extend our analysis and examine other open quantum system models, such as the
more complex spin-boson (SB) model,  
$\hat H = \frac{\Delta}{2}\hat\sigma_z + \hat\sigma_x \sum_k f_k (\hat c_k^{\dagger} +\hat c_k) + \sum_k \nu_k \hat c_k^{\dagger}\hat c_k$.
Unlike the pure decoherence model, this model displays both decoherence and energy relaxation dynamics.

The dynamics of the SB model was examined in many works with a range of techniques \cite{Weiss}.
Particularly, it was simulated in Ref. \cite{Nazir16} by benchmarking
the RC-QME method to  numerically exact simulations,  
manifesting an excellent agreement even at strong coupling.
Quantifying non-Markovian dynamics in the SB dynamics (or other non-integrable models) based on the BLP and the RHP measures, or others, is a highly nontrivial task, given that the model does not have a closed-form solution and that it is challenging to simulate it in a numerically-exact manner.
Simulations of the trace-distance were  performed e.g. in Ref. \cite{Thorwart} based on a real-time path integral method and more recently
 in Ref. \cite{Thoss} using the multilayer multiconfiguration time-dependent Hartree approach (ML-MCTDH). 
Perturbative approaches for dynamics were employed e.g. in Refs. \cite{Breuer12,Nori15,Rivas17,Vega17}.

Given the complexity of the problem, we now simply demonstrate with the RC-QME method 
that the SB model at small $\gamma$ shows recoherence dynamics, see Fig. \ref{fig:SB_coherence}(c)-(d).
These recoherences do not show up
with the BMR-QME method, which is Markovian by construction.
As such, we argue that these features indicate on non-Markovian dynamics.
We reiterate that the RC-QME method should be quite accurate in the small $\gamma$ regime,
and even at strong coupling \cite{Nazir16}.
More details on the SB model and the corresponding dynamics of spin polarization, are 
discussed in Appendix C.

Future work will be dedicated to capturing and quantifying strong coupling and non-Markovian effects with the RC-QME in different open quantum system models. For example, the equilibrium behavior of the
V-model was recently examined in Ref. \cite{Brumer-rev, Cresser} at weak---and ultra strong coupling. It remains an open topic to explore the dissipative dynamics of such multi-level models at intermediate and strong coupling.

Proposals for novel quantum devices rely on nontrivial aspects of small, quantum systems \cite{Janet-rev}.
Specifically, strong system-bath couplings and bath structuring are suggested as means to realize new effects. 
The RC-QME method has been proving itself as a powerful tool in this respect. 
It nicely complements numerically exact approaches, 
which typically struggle to converge in regimes that are favorable for RC-QME simulations. 
Future work will be focused on studies of quantum thermal machines beyond weak system-bath coupling,
towards the discovery of their unique performance. 

\begin{acknowledgements}
DS acknowledges support from an NSERC Discovery Grant and the Canada Research Chair program.
\end{acknowledgements}


\section*{Appendix A: Markovian Limit of the pure decoherence model}
\renewcommand{\theequation}{A\arabic{equation}}
\renewcommand{\thesection}{A\arabic{section}}
\setcounter{equation}{0}  
\setcounter{section}{0}
\label{app:1}

In this Appendix, we derive an expression for 
the decoherence timescale $\tau_D$ in the pure decoherence model.
We perform our analysis on the model in its original representation, Eq. (\ref{eq:Hamiltonian_Dephasing}) using
the Brownian spectral function for the bath, Eq. (\ref{eq:Brownian}).

First, we derive the decoherence timescale by simplifying the exact analytic expression. 
Next, we show that one can reach the same result from the Redfield equation of motion, a method that relies on the Born-Markov approximation. 
%

\subsection{Derivation based on the exact analytic expression}
\label{sec:A1}

We study 
the long time limit of the decoherence function \cite{Breuer}, 
\bea
\Gamma(t) = -4\int_0^\infty d\omega J(\omega) \coth(\frac{\beta \omega}{2}) \frac{1 - \cos(\omega t)}{\omega^2}.
\eea
Unlike textbook calculations, which simplify this function assuming an Ohmic spectral function \cite{Breuer},
here we use a Brownian spectral function for the bath,
\bea
J(\omega) = \frac{4 \omega \Omega^2 \lambda^2 \gamma/\pi}{(\omega^2 - \Omega^2)^2 + (2 \gamma \Omega \omega)^2}.
\label{eq:JA}
\eea
Noting that $1 - \cos(\omega t) = 2\sin^2\left(\frac{\omega t}{2}\right)$
and that in the long time limit 
$\frac{4 \sin^2(\frac{\omega t}{2})}{\omega^2} \to 2\pi t \delta(\omega)$,
we get
\bea
    \Gamma(t) \to -2 \pi t \left[J(\omega) \coth(\frac{\beta \omega}{2})\right]\Bigg|_{\omega\to 0}.
\eea
%
For nonzero temperatures, 
$\left[J(\omega) \coth(\frac{\beta \omega}{2})\right]\Big|_{\omega\to 0}\to \frac{8\gamma \lambda^2}{\pi \Omega^2\beta}$,
thus we obtain in the long time limit,
\bea
    \Gamma(t) = -\frac{16 \gamma \lambda^2}{\Omega^2\beta} t.
\eea
According to this result, coherences decay exponentially in time with the
decoherence timescale
\bea
    \tau_D = \frac{\Omega^2 \beta}{16 \gamma \lambda^2}.
\eea
We can now be precise and note that the ``long time limit" corresponds 
to $t\gg \tau_D$. Thus, the exponential form 
is arrived at more quickly as we increase the temperature and the coupling strength $\lambda$. 




\subsection{Derivation from the Born-Markov Redfield Equation}
\label{sec:A2}

In the Schr\"odinger representation, the Redfield QME is given by
\bea
&&\dot \rho(t)  = -i[\hat H_S,\rho(t)]
\nonumber\\
&&\hspace{10mm}- \int_0^{\infty} d\tau
\Big \{ [\hat S,e^{-i\hat H_S\tau} \hat S e^{i\hat H_S\tau}  \rho(t)] \langle \hat B(t) \hat B(t-\tau)
 \rangle
\nonumber\\
&&\hspace{20mm}- [ \hat S, \rho(t) e^{-i\hat H_S\tau}\hat S e^{i\hat H_S\tau}] \langle \hat B(t-\tau) \hat B(t) \rangle
\Big\}.
\nonumber\\
\eea
Here, $\hat S$ is a system operator that couples to the bath operator, $\hat B$, with $\hat B(t)$ an interaction 
representation operator.
%
In our model, the system's Hamiltonian is $\hat H_S = \frac{\Delta}{2}\hat \sigma_z$ and
$\hat S =\hat \sigma_z$; $\hat \sigma_z = |1\rangle \langle 1| -|0\rangle \langle 0|$.
Since we are interested in the dynamics of coherences,
we focus on the matrix element $\rho_{01}(t) = \bra{0}\rho(t)\ket{1}$. 
The above expression simplifies to
\bea
\dot{\rho}_{01}(t) = -i\Delta \rho_{01}(t) 
- 2\rho_{01}(t) \int_{-\infty}^{\infty} d\tau \langle \hat B(\tau)\hat B(0) \rangle,
\label{eq:rhotA}
\eea
where we assumed that the bath is stationary thus
$\langle \hat B(0) \hat B(\tau) \rangle = \langle \hat B(-\tau) \hat B(0)\rangle$, allowing us
to combine two bath correlation functions into a single integral. 
To compute the bath correlation function we note that 
$\hat B(\tau) = \sum_k f_k (\hat b_k e^{-i\nu_k \tau}+ \hat b^{\dagger}_k e^{i\nu_k \tau})$. 
Using this and moving to the continuum limit we get
\bea
 \langle \hat B(\tau) \hat B(0)\rangle &=& \int_0^{\infty} d\omega J(\omega) \left\{ 
\left[1 + n_B(\omega)\right]e^{-i\omega \tau} + n_B(\omega)e^{i\omega \tau} \right\},
    \nonumber \\
    &&
\eea
with $n_B(\omega) = (e^{\beta \omega} - 1)^{-1}$ the Bose-Einstein distribution function.
It is now possible to perform the time integral in Eq. (\ref{eq:rhotA})
using $2\pi\delta(\omega) = \int_{-\infty}^{\infty} d\tau e^{i\omega \tau}$,
\bea
    \dot{\rho}_{01}(t) &=& -i\Delta \rho_{01}(t) 
- 4\pi\rho_{01}(t)\int_{0}^{\infty} d\omega J(\omega) [2n_B(\omega) + 1]\delta(\omega). 
    \nonumber \\
    &&
\eea
Evaluating the trivial integral we get
\bea
\dot{\rho}_{01}(t) = 
-i\Delta \rho_{01}(t)  - 2\pi J(0)\left[2n_B(0) + 1\right] \rho_{01}(t).
\label{eq:AppAdiff}
\eea
Care needs to be taken when evaluating the second term as $n_B(0)$ is divergent. 
To keep the discussion general for now we define $A \equiv 2\pi \lim_{\omega\to\ 0} J(\omega)[2n_B(\omega) + 1]$,
assuming this expression has a well-defined limit.
The inverse of this term corresponds to the decoherence timescale, $\tau_D = A^{-1}$.
Altogether, the differential equation Eq. (\ref{eq:AppAdiff}) solves to
\bea
    \rho_{01}(t) = \rho_{01}(0)e^{-i\Delta t} e^{-At}.
\eea
As expected, the Born-Markov approximation results in an exponentially decaying dynamics for $\rho_{01}$.
Coherences are affected by two timescales. 
While $\Delta^{-1}$ is related to intrinsic coherent 
oscillations, $\tau_D$ describes the loss of coherence with time due to the coupling to
the bath.
%
We can now explicitly evaluate this latter timescale using the Brownian spectral function, 
resulting in
 $   \tau_D = \frac{\beta \Omega^2}{16 \gamma \lambda^2}$,
which agrees with results from Sec. A of this Appendix.


\section*{Appendix B: Details on the Redfield equation}
\renewcommand{\theequation}{B\arabic{equation}}
\renewcommand{\thesection}{B\arabic{section}}
\setcounter{equation}{0}  
\setcounter{section}{0}
\label{app:2}

For completeness, we provide here the full reaction-coordinate Redfield equation that we use in the main text to simulate the dynamics of the supersystem.
Working in the Schr\"odinger picture 
and in the energy-basis of the subsystem, 
the Redfield equation for the reduced density matrix of the supersystem (dimension 2$M$) 
is given by
\bea
    \dot \rho_{ES,mn}(t) &=& -i\omega_{mn} \rho_{ES,mn}(t)
     \nonumber\\
    &+&\sum_{j,k} \Big[ R_{mj,jk}(\omega_{kj}) \rho_{ES,kn}(t) + R^*_{nk,kj}(\omega_{jk})\rho_{ES,mj}(t)
     \nonumber\\
    &-& R_{kn,mj}(\omega_{jm}) \rho_{ES,jk}(t) - R^*_{jn,mk} \rho_{ES,jk}(t) \Big].
\eea
Here $\omega_{mn}$ are the Bohr frequencies 
of the extended system.  
The dissipator itself is given by half Fourier transforms of the bath autocorrelation function,
\bea
    R_{mn,jk}(\omega) &=& (S_{ES}^D)_{mn} (S_{ES}^D)_{jk} \int_0^{\infty} d\tau e^{i\omega \tau} \langle \hat B(\tau) \hat B(0) \rangle
    \\ \nonumber
    &=& (S_{ES}^D)_{mn} (S_{ES}^D)_{jk} [\Gamma_{RC}(\omega) + i\Delta_{RC}(\omega)].
\eea
In simulations, we neglect the imaginary part of the Redfield tensor,  $\Delta_{RC} (\omega)$.
Detailed discussions about the role of this self energy 
in the steady state limit for specialized models (beyond the
decoherence model and the spin-boson model discussed in this work)
appear in \cite{Archak, Cresser}. 
%
The real portion of the correlation function can be readily evaluated to yield
\bea
\Gamma_{RC}(\omega) =
\begin{cases}
\pi J_{RC}(\omega) n(\omega) & \omega > 0   \\
\pi J_{RC}(|\omega|)[(n(|\omega|) + 1] & \omega < 0,\\
\gamma/\beta & \omega = 0 \,\,\,({\rm Ohmic\,\, model}).
\end{cases}
\eea
 $n(\omega)$ is the Bose-Einstein distribution function and $J_{RC}(\omega)$  is
the spectral function of the residual bath given by Eq. (\ref{eq:Ohmic}).

\section*{Appendix C: Non-Markovian effects in the spin-boson model}
\renewcommand{\theequation}{C\arabic{equation}}
\renewcommand{\thesection}{C\arabic{section}}
\setcounter{equation}{0}  
\setcounter{section}{0}
\label{app:3}

\begin{figure}
  \centering
    \includegraphics[width=\columnwidth]{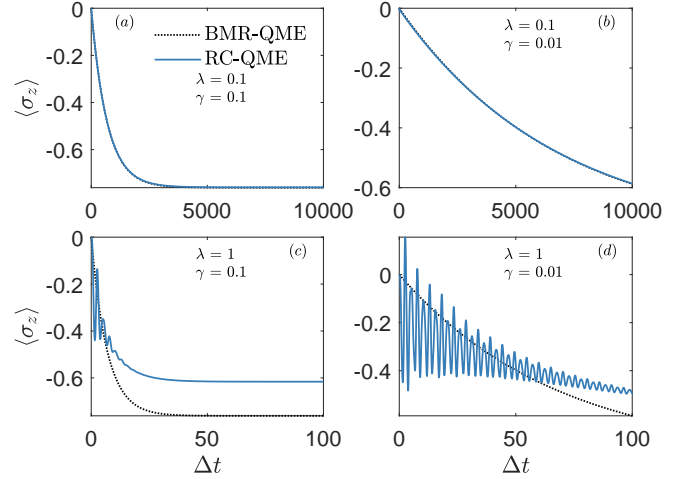}
    \caption{Polarization dynamics in the spin-boson model, $\langle \sigma_z(t)\rangle$.
with the initial  state $\rho = |-\rangle\langle -|$
using the BMR-QME method (dotted line) and the RC-QME method (full).
Parameters are the same as in Fig. \ref{fig:Coherence}.
}
    \label{fig:SB_Population}
\end{figure}

The decoherence model, Eq. (\ref{eq:Hamiltonian_Dephasing}),
does not allow dissipative (energy exchange) dynamics.
In contrast, the spin-boson model
\bea
   \hat H = \frac{\Delta}{2}\hat\sigma_z + \hat\sigma_x \sum_k f_k (\hat c_k^{\dagger} +\hat c_k) + \sum_k \nu_k \hat c_k^{\dagger}\hat c_k
    \label{eq:Hamiltonian_Spin_Boson}
\eea
displays more complex dissipative and decoherence dynamics.
Similarly to Eq. (\ref{eq:Hamiltonian_Dephasing}), 
$\Delta$ is the spin splitting, $\hat \sigma_z = |1\rangle \langle 1| -|0\rangle \langle 0|$,
$\hat \sigma_x= |1\rangle \langle 0| +|0\rangle \langle 1|$,
$\hat c_k^{\dagger}$ ($\hat c_k$) are bosonic creation (annihilation) operators for mode 
$k$, with $f_k$ a coupling energy.
While the dynamics of the SB model can be analytically described in different limits \cite{Weiss},
a complete, exact analytic solution is unavailable.

Applying the reaction coordinate mapping to the spin-boson model,
Eq. (\ref{eq:Hamiltonian_Spin_Boson}), we get the following Hamiltonian
\bea
\hat H_{RC} &=& \frac{\Delta}{2}\hat \sigma_z + \lambda \hat \sigma_x (\hat a^{\dagger} + \hat a) + \Omega \hat a^{\dagger}\hat a  +  (\hat a^{\dagger} + \hat a)^2\sum_k\frac{g^2_{k}}{\omega_{k}}
\nonumber\\
&+&  (\hat a^{\dagger}+\hat a)\sum_k g_{k} (\hat b_{k}^{\dagger}+\hat b_{k}) + \sum_{k}\omega_{k}\hat b_{k}^{\dagger}\hat b_{k}.
\label{eq:RC_Spin_Boson_Hamiltonian}
\eea
%
%
The original model takes a Brownian spectral function for the bath. After the RC mapping, the residual bath is coupled (presumably) weakly to the system with an Ohmic spectral function.

We now focus on the appearance of non-Markovian dynamics as we reduce $\gamma$ and increase $\lambda$.
Since quantifying non-Markovian dynamics using the BLP and RHP measures
is  nontrivial for the spin-boson model, 
we only examine this concept qualitatively
by comparing the RC-QME results to Markovian simulations.
The dynamics of coherences in the spin-boson model is displayed in the main text, Fig.  \ref{fig:SB_coherence}.
There, we show that
as we increase the coupling strength and reduce $\gamma$ we
see pronounced deviations from Markovian dynamics, captured by the RC method.
The corresponding dynamics of polarization is depicted in Fig.  \ref{fig:SB_Population}.
We find that the BMR-QME method misses the damped coherent oscillations in the dynamics 
that show up at strong coupling and small $\gamma$.
Furthermore, the BMR-QME method no longer provides the correct long-time solution at strong coupling.

The RC-QME method predicts significant deviations form
Markovian dynamics in the spin-boson model once we structure the bath and enhance coupling strength.
Benchmarking RC-QME results at small $\gamma$  against numerically exact simulations is challenging:
For example,  consider the   quasi-adiabatic propagator path integral (QuAPI)
method \cite{Makri1,Makri2}:
For $\gamma=0.01$ and $\Omega=3$,  the bath memory time may be approximated by 
$\tau_M\sim 1/(\gamma\Omega)\sim 30$.
To properly capture the damped coherent oscillations and reduce the Trotter error
 (of period $2\pi/\Delta$) 
one needs to adopt a short time step of $\delta t \sim 0.2$.
Thus, to cover the memory time one needs a memory kernel with $N\sim\tau_M/\delta t\sim 200$
segments.
This significant task may be achieved with advanced implementations of numerically exact path integral
approaches \cite{Makri20a,Makri20b,Keeling}, and we leave it to future work.


\section*{Appendix D: Additional Convergence Results}
\renewcommand{\theequation}{D\arabic{equation}}
\renewcommand{\thesection}{D\arabic{section}}
\setcounter{equation}{0}  
\setcounter{section}{0}
\label{app:4}

\begin{figure} [ht]
\vspace{5mm}
\includegraphics[width=\columnwidth]{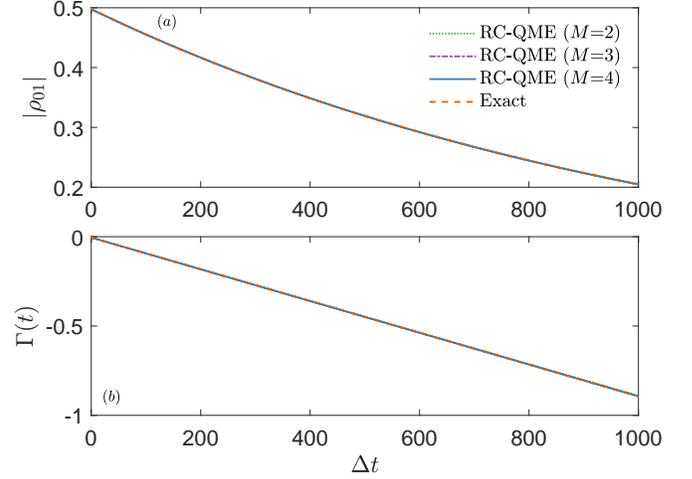}
\caption{Convergence of RC-QME simulations, Fig. \ref{fig:Coherence}(a),
with respect to $M$ at weak coupling, $\lambda=0.1$.
We present (a) the decoherence dynamics and (b) corresponding decoherence function for $M=2$ (dotted),
3 (dashed-dotted), 4 (full), and compare those to exact results (dashed).
Parameters are $\lambda=0.1$ and $\gamma=0.1$, as well as
$\Delta = 1$, $\Omega = 3\Delta$, $T = 0.5 \Delta$, $\Lambda = 1000 \pi \Delta$,
as in Fig. \ref{fig:Coherence}(a).
}
\label{fig:Convergence_lambda_01}
\end{figure}

\begin{figure}[ht]
\vspace{10mm}
\includegraphics[width=\columnwidth]{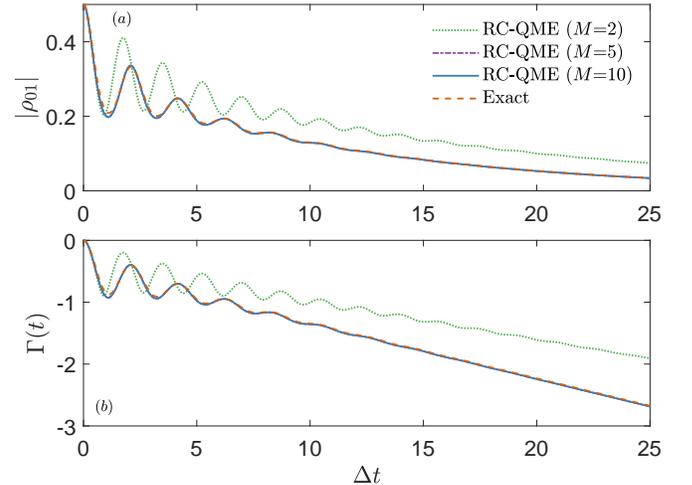}
\caption{Convergence of RC-QME simulations, Fig. \ref{fig:Coherence}(c),
with respect to $M$ at intermediate coupling, $\lambda=1$.
We present (a) the decoherence dynamics and (b) corresponding
decoherence function for $M=2$ (dotted), 5 (dashed-dotted), 10 (full),
and compare those to exact results (dashed).
Parameters are $\lambda=1$ and $\gamma=0.1$ as well as
$\Delta = 1$, $\Omega = 3\Delta$, $T = 0.5 \Delta$, $\Lambda = 1000 \pi \Delta$
as in Fig. \ref{fig:Coherence}(c).
}
\label{fig:Convergence_lambda_1}
\end{figure}

The computational complexity of the RC-QME method scales as $O((d\times M)^4)$ 
due to the need to construct the Redfield tensor for the 
$(d  M)\times(d M)$ reduced density matrix.
Here, $d$ is the dimension of the original subsystem (2 for spin) 
and $M$ the number of truncated levels in the RC.

In this Appendix, we present additional $M$-convergence results,
analogous to Fig. \ref{fig:Convergence_lambda_5}. 
There, we focused on the challenge in 
converging coherences once the system-bath coupling strength was made large,
$\lambda > \Omega$.
In contrast, here in Figs. \ref{fig:Convergence_lambda_01} and \ref{fig:Convergence_lambda_1} 
we show that in both weak and intermediate couplings, $\lambda < \Omega$, convergence is simple, 
pointing to the utility of the RC-QME method.

First, at weak coupling we show in Fig. \ref{fig:Convergence_lambda_01} that 
even just including $M=2$ levels in the RC is already sufficient for results to be converged. 
The intermediate coupling regime is presented in
Fig. \ref{fig:Convergence_lambda_1}; here convergence with respect to $M$ is 
reached at $M = 5$, which is still a simple computational task.
Altogether, we emphasize that
ensuring convergence with respect to $M$ is a critical  aspect of the RC-QME technique.
In the pure-decoherence model presented in this study 
convergence poses a technical challenge only in the strong coupling regime.

\end{document}